\documentclass[prd,11pt,nobibnotes,nofootinbib]{revtex4}
\usepackage[a4paper,scale=0.73]{geometry}
\pdfoutput=1

\let\Oldbar\bar
\usepackage{amsfonts,amssymb,amsmath}
\let\bar\Oldbar

\usepackage{amsmath,amssymb}
\usepackage{graphicx}
\usepackage[T1]{fontenc}
\usepackage{lmodern}

\usepackage{epstopdf}

\newcommand{\longpage}[1][1]{}
\newcommand{\shortpage}[1][1]{}

\DeclareMathOperator{\Tr}{Tr}%

\newcommand{\expp}[1]{ \mathop\mathit{e}\nolimits^{#1}}

\newcommand{\derp}[3][]{\frac{\partial^{#1} #2}{\partial #3^{#1}}}

\newcommand{\ud}[2][]{\textrm{d}^{#1}{#2}\,}

\newcommand{\vd}[2][]{\textrm{d}^{#1}{#2}}
\newcommand{\Eqref}[1]{Eq.~\eqref{#1}}
\newcommand{\ie}{\emph{i.e.}}

\renewcommand{\Re}{\mathop\mathrm{Re}\nolimits}
\renewcommand{\Im}{\mathop\mathrm{Im}\nolimits}
\newcommand{\vect}{\mathbf}

\newcommand{\udpi}[2][]{\frac{\textrm{d}^{#1}{#2}}{(2\pi)^{#1}}}

\newcommand{\GF}{G_\mathrm{F}}

\newcommand{\SigmaR}{\Sigma_\mathrm R}

\newcommand{\Gret}{{G}_\mathrm R}

\newcommand{\phim}{\phi_{m}}
\newcommand{\phiM}{\phi_{M}}

\newcommand{\dm}{\mathit{\Delta m}}
\newcommand{\GR}{\Gret}

\newcommand{\SigmaN}{\Sigma^{(1)}}

\newcommand{\GN}{{G^{(1)}}}
\newcommand{\GA}{G_\text{A}}
\newcommand{\pplus}{{\scriptscriptstyle (+)}}

\newcommand{\SigmaImp}{\widetilde\Sigma_\text{R}}

\allowdisplaybreaks[3]

\providecommand{\citep}{\cite}

\newcommand{\ret}{}

\begin{document}
\author{Daniel Arteaga}
\email{darteaga@ub.edu}
\affiliation{Departament de F\'\i sica Fonamental and Institut de Ci\`encies del Cosmos, Facultat de F\'\i sica, Universitat de Barcelona, Av.~Diagonal 647, 08028 Barcelona (Spain)}

\title{A field theory characterization of interacting adiabatic particles in cosmology}

\begin{abstract}
We explore the adiabatic particle excitations of an interacting field in a cosmological background. By following the time-evolution of the quantum state corresponding to the particle excitation, we show how the basic properties characterizing the particle propagation can be recovered from the two-point propagators. 
As an application, we study the background-induced dissipative effects  on the propagation of a two-level atom in an expanding universe.
\end{abstract}

\maketitle

\section{Introduction}

Much attention has been given in the literature to the different particle concepts existing in curved spacetime \cite{BirrellDavies,WaldQFT,Fulling}, which can be roughly classified into three classes. First, there are the global particle concepts which depend on special properties of the spacetime: if asymptotically flat regions exist, particles can be defined with respect to those asymptotic observers; if the spacetime has enough symmetries, particles can be defined with respect to them. Second, whenever there is   a separation of scales between the propagating degrees of freedom and the background,  particles can be defined in full generality, without making any reference to the specific form of the spacetime:  these are the adiabatic and quasilocal particle concepts. Finally, the less restrictive particle concept is  the operational definition in terms of the response of a quantum mechanical detector.

Most analysis have however dealt with non-interacting particles, \ie, with particles which are free except for the classical gravitational interaction with the curved background. When considering interacting processes in the universe, as for instance in nucleosyntehesis, the curved background can be often safely neglected. However, there are situations in which both the quantum interaction between the particles and the classical gravitational interaction with the background spacetime can be relevant: for instance it has been suggested \cite{tHooft96,Parentani01a,Parentani01b,Parentani02,tHooft06} that this might be the case  when considering the trans-Planckian problem \cite{Jacobson91,Jacobson93,Jacobson99} both in black hole physics and cosmology.

While there are several works studying interacting particles in curved spacetimes (see for instance \cite{BirrellDavies,Ford83,AudretschSpangehl85,AudretschSpangehl86,Audretsch86,AudretschSpangehl87,CespedesCampos90,CespedesCampos92,Tsaregorodtsev95,BrosEtAl07} and references therein), most of them consider global particle concepts which rely on the existence of asymptotic regions or on specific symmetry properties of the spacetime. However, other particle interpretations which do not require particular properties of the spacetime can also be extended to the interacting case. In particular, when scales are well separated in cosmology, it is natural to look for an extension of the adiabatic particle concept. Indeed interacting adiabatic particles were already considered in Refs.~\cite{ArteagaParentaniVerdaguer07,Arteaga07a}, which dealt with the disspative effects on the propagation a two-level atom in a an expanding universe.

In this paper we pursue the work of Refs.~\cite{ArteagaParentaniVerdaguer07,Arteaga07a} in two different directions. On the one  hand we establish a general framework for the analysis of adiabatic interacting particles and quasiparticles in cosmology, thereby extending the results of Ref.~\cite{Arteaga08a} to cosmological backgrounds. We will construct and follow the time evolution of the quantum state corresponding to the (quasi)particle excitations, investigating how the two-point functions can be connected to observable quantities. 
The propagation of a quasiparticle in a physical medium and the propagation of a particle in a curved background have many similarities \cite{ArteagaParentaniVerdaguer04a}, and both situations can be treated using similar techniques. In particular, the closed time path  (CTP) method \cite{Schwinger61,Keldysh65,ChouEtAl85,CalzettaHu88,CamposVerdaguer96,Weinberg05} provides a generic way to deal with field theories over arbitrary spacetime backgrounds with an arbitrary field states. (For a brief introduction to the CTP method adapted to the notation of this paper see appendix A in Ref.~\cite{Arteaga07a}.) 

On the other hand, using this framework, we concentrate on the novel dissipative effects induced by the universe expansion, beyond those present in flat spacetime. We expect those effects to be relevant when the interaction timescale is of the order of the expansion timescale. Indeed, in Ref.~\cite{Arteaga07a} it was mentioned  that within this formalism one can readily show that the two-level atom becomes excited when propagating in a de Sitter spacetime in the presence of a conformal radiation field in the vacuum, the excitation rate corresponding to the effective de Sitter temperature \cite{BirrellDavies}. In this paper we elaborate on this point by making explicit  the calculation and generalizing the result to other backgrounds.

It will be important to treat appropriately the several timescales appearing in the problem: the propagation timescale $E_{\vect k}^{-1}$, corresponding to the inverse de Broglie frequency of the particle; the interaction timescale $t_\text{int}$; the universe expansion timescale $H^{-1}$, given by the inverse Hubble rate, and finally the observation timescale $t_\text{obs}$. Two timescale hierarchies will be assumed thought the paper. First, in order for the adiabatic particle concept to be applicable, the typical propagation time $E_{\vect k}^{-1}$ must be much shorter than the typical expansion time of the universe $H^{-1}$,  so that one can  look for an analytic Wentzel-Kramers-Brillouin (WKB) approximation for the two-point propagators. Second, the observation timescale $t_\text{obs}$ will be assumed to be always much larger than the interaction timescale $t_\text{int}$, so that the asymptotic field theory results can be (approximately) applied. Depending on the other timescale hierarchies several cases will be discussed in the paper.

The paper is organized as follows. In Sect.~II we present the relation between the propagators and self-energies in curved spacetime, making the connection with the CTP formalism. In Sect.~III we give a field-theoretic description of interacting adiabatic particles and quasiparticles in cosmological backgrounds by following the time-evolution of the expectation value of the Hamiltonian operator.  In Sect.~IV we apply the results to study the dissipative effects in the propagation of a two-level atom. Finally, in Sect.~V we summarize and discuss the main results. 

Throughout the paper we use a signature $(-,+,+,+)$ and a system of natural units with $\hbar=c=1$. The same symbol is used for a quantity and its Fourier transform provided there is no danger of confusion.

\section{Interacting fields in curved backgrounds}

In the following we consider the general situation in which there is a  scalar field in a generic state $\hat\rho$, over a globally hyperbolic spacetime characterized by some metric $g_{\mu\nu}$.
Most work studying interacting fields over general curved backgrounds was developed during the late 70s and early 80s, and was focused on the study of the renormalizability of the theories (see Ref.~\cite{BirrellDavies} and references therein). More recent works have focused on general properties of interacting fields in curved backgrounds \cite{FriedmanEtAl92,HollandsWald01,HollandsWald02,HollandsWald03,HollandsWald05} and on the particular case of de Sitter and anti-de Sitter  \cite{BrosEtAl94,BrosMoschella96,EinhornLarsen03,JoungEtAl06,JoungEtAl07}. In this paper we  will not attempt to make any review of the subject, nor make a complete presentation of interacting quantum fields in curved backgrounds; we will simply highlight some aspects relevant for us.

The Feynman propagator, positive and negative Wightman functions and Dyson propagator,
\index{Propagator!Feynman}
\index{Propagator!Wightman}
\index{Propagator!Dyson}
\begin{subequations} \label{CorrFunct}
\begin{align}
    G_{11}(x,x') &= G_\text{F}(x,x') := \Tr{\big(
\hat\rho\, T \hat\phi(x) \hat\phi(x') \big)
}, \\
    G_{12}(x,x') &= G_{+}(x,x') := \Tr{\big(
\hat\rho\,  \hat\phi(x) \hat\phi(x') \big)
}, \\
    G_{21}(x,x') &=G_{-}(x,x') := \Tr{\big(
\hat\rho\,  \hat\phi(x') \hat\phi(x) \big)
}, \\
    G_{22}(x,x') &=G_\text{D}(x,x') := \Tr{\big(
\hat\rho\,  \widetilde T \hat\phi(x) \hat\phi(x')\big)
 },
\end{align}
\end{subequations}
respectively, are  correlation functions appearing in the CTP formalism  (which is natural when dealing with interaction in curved spacetimes), and can be conveniently organized in a $2\times 2$ matrix $G_{ab}$:
\begin{equation}
	G_{ab}(x,x') = 
		\begin{pmatrix}
			\GF(x,x') & G_-(x,x') \\
			G_+(x,x') & G_\text{D}(x,x')
 		\end{pmatrix}.
\end{equation}
We may also consider the Pauli-Jordan or commutator propagator,
\index{Propagator!Pauli-Jordan}
\begin{subequations}\label{CorrFunct2}
\begin{equation}
 G(x,x') := \Tr{\big(
\hat\rho\,[  \hat\phi(x) ,\hat\phi(x')] \big)
},
\end{equation}
and the Hadamard or anticonmutator function
\index{Propagator!Hadamard}
\begin{equation}
	G^{(1)}(x,x') := \Tr{\big(
\hat\rho\,\{  \hat\phi(x) ,\hat\phi(x') \} \big)}.
\end{equation}
\end{subequations}
Finally, one can also consider the retarded and advanced propagators,
\index{Propagator!retarded}
\index{Propagator!advanced}
\begin{subequations}\label{CorrFunct3}
\begin{align}
    G_\mathrm {R}(x,x') &:= \theta(x^0-x'^0) G(x,x') =\theta(x^0-x'^0) \Tr{\big(
\hat\rho\, [\hat \phi(x),\hat\phi(x')]\big)}, \label{RetardedProp}\\
    G_\mathrm {A}(x,x') &:= \theta(x'^0-x^0) G(x,x') =\theta(x'^0-x^0) \Tr{\big(
\hat\rho\, [\hat \phi(x),\hat\phi(x')]\big)},
\end{align}
\end{subequations}

\index{Schwinger-Dyson equation}
Self-energies $\Sigma^{ab}(x,x')$ are introduced through the Schwinger-Dyson equation:
\begin{equation} \label{SelfEnergyGeneralCurved}
\begin{split}
    G_{ab}(x,x') = G_{ab}^{(0)}(x,x') &+
    \int {\vd[4]{z}}\sqrt{-g(z)}\, \vd[4]{z'}\sqrt{-g(z')}  G_{ac}^{(0)}(x,z) [-i\Sigma^{cd}(z,z')]  G_{db}(z',x'),
\end{split}
\end{equation}
where $G^{(0)}_{ab}(x,x')$ are the propagators of the corresponding free theory and an Einstein summation convention has been assumed for the repeated CTP indices $a,b,c\ldots\in\{1,2\}$. The self-energy components $\Sigma_{ab}(x,x')$ can be computed as the sum of all one-particle irreducible diagrams beginning in a vertex type $a$ and ending in a vertex type $b$. 

It will be useful to work with rescaled fields, propagators and self-energies as follows:
\begin{subequations}\label{rescaling}
	\begin{align}
		\bar\phi(x) &:= [-g(x)]^{1/4}, \\
		\bar G_{ab}(x,x') &:= [-g(x)]^{1/4}  G_{ab}(x,x') [-g(x')]^{1/4}, \\
		\bar \Sigma_{ab}(x,x') &:= [-g(x)]^{1/4}  \Sigma_{ab}(x,x') [-g(x')]^{1/4}.
	\end{align}
\end{subequations}
With these definitions $\bar\phi(x)$, $\bar G_{ab}(x,x')$ and  $\bar \Sigma_{ab}(x,x')$ are (bi)scalar densities of weight 1/2. In terms of the bar quantities, the relation between propagators and self-energies becomes identical to flat spacetime:
\begin{equation} \label{SelfEnergyGeneralCurved2}
\begin{split}
    \bar G_{ab}(x,x') = \bar G_{ab}^{(0)}(x,x') &+
    \int {\vd[4]{z}}\, \vd[4]{z'} \ \bar G_{ac}^{(0)}(x,z) [-i\bar\Sigma^{cd}(z,z')]  \bar G_{db}(z',x').
\end{split}
\end{equation}

The retarded propagator obeys a direct relation with the retarded self-energy $\Sigma_\mathrm R(x,x') := \Sigma^{11}(x,x') +
\Sigma^{12}(x,x')$,
\begin{equation} \label{SelfEnergyGeneralRetardedCurved}
    \bar\GR(x,x') = \bar\GR^{(0)}(x,x')+
    \int \ud[4]{z} \ud[4]{z'} \bar\GR^{(0)}(x,z) [-i\bar\SigmaR(z,z')]  \bar\GR(z',x').
\end{equation}
A similar relation holds between the advanced propagator and advanced self-energy. Another useful combination is the Hadamard self-energy, which is defined as $\Sigma^{(1)}(x,x') = \Sigma^{11}(x,x') + \Sigma^{22}(x,x')$ [or equivalently as $\Sigma^{(1)}(x,x') =- \Sigma^{12}(x,x') - \Sigma^{21}(x,x')$]   and which is
 related to the Hadamard propagator through\footnote{\Eqref{GSigmaN} is not valid in full generality, but requires remote initial conditions and a sufficiently dissipative behavior. See Ref.~\cite{ArteagaThesis} for further details.} 
 \begin{equation} \label{GSigmaN}
	\bar G^{(1)}(x,x') = -i \int \ud[4]{y} \ud[4]{y'} \bar\GR(x,y) \bar\SigmaN(y,y') \bar\GA(y',x').
\end{equation}

In a spatially flat Friedmann-Lema\^{\i}tre-Robertson-Walker background, with metric
\begin{equation}
	\vd s ^2 = -\vd t^2 + a^2(t) (\vd x^2+\vd y^2 + \vd z^2) 	
\end{equation}
the field operator can be expanded in conformal modes, $\phi(t,\vect x) = \sum_\vect k\phi_\vect k(t) \expp{i \vect k \cdot \vect x}$. In practice this means that each conformal mode can be analyzed separately and that we can exploit conservation of the conformal momentum.
Eq.~\eqref{SelfEnergyGeneralCurved} can be particularized to the cosmological case:
\begin{equation}\label{DysonCosmo}
	\bar G_{ab}(t,t';\vect k) = \bar G_{ab}^{(0)}(t,t';\vect k) -i 
	\int \ud s \ud {s'}  \bar G_{ac}^{(0)}(t,s;\vect k) 
	\bar\Sigma^{cd}(s,s';\vect k) \bar G_{db}(s',t';\vect k),
\end{equation}
where the rescaled fields are $\bar \phi_\vect k(t) = a^{3/2}(t) \phi_\vect k(t)$. Two particular relations will be of interest: the one corresponding to the retarded propagator,
\begin{subequations}
\begin{equation}\label{Dyson}
	\bar \GR(t,t';\vect k) = \GR^{(0)}(t,t';\vect k) -i 
	\int \ud s \ud {s'}  \bar \GR^{(0)}(t,s;\vect k) 
	\bar\SigmaR(s,s';\vect k) \bar \GR(s',t';\vect k),
\end{equation}
and the one corresponding to the Hadamard function:
\begin{equation} \label{SelfEnergyHadamard2}
	\bar \GN(t,t';\vect k) = -i 
	\int \ud s \ud {s'}  \bar \GR(t,s;\vect k) 
	\bar\SigmaN(s,s';\vect k) \bar \GA(s',t';\vect k).
\end{equation}
\end{subequations}

In expanding universes the propagators, though space-translation invariant, are not
time-translation invariant. 
Nevertheless, we can always 
express the propagator 
in a
Fourier transform 
with respect to the difference variable $\Delta=t-t'$, 
while keeping  $T=(t+t')/2$ constant:
\begin{equation} \label{mixed}
	\bar\GR(\omega,T;\vect k):=\int \ud \Delta \expp{i\omega\Delta} 
	\bar\GR(T+\Delta/2,T-\Delta/2;\vect k) \, .
\end{equation}
In general, the Fourier transform does not help in simplifying the equations any further. However, if both typical propagation and interaction times are much shorter than the typical expansion time, \ie\ if the hierarchy
$
	E_\vect k^{-1}, t_\text{int} \ll t_\text{obs} \ll H^{-1}.
$ holds, the retarded propagator can be solved from \Eqref{Dyson}:
\begin{equation} \label{shortime}
	\bar\GR(\omega,T;\vect k) \approx \frac{ -i}
	{[-i \bar \GR^{(0)}(\omega,T;\vect k)]^{-1} + 
	\bar\SigmaR(\omega,T;\vect k)  }\, ,
\end{equation}
where the
free propagator is approximated by
$
	 [-i \bar \GR^{(0)}(\omega,T;\vect k)]^{-1} \approx - \omega^2 + {\vect k^2 }/{a^2(T)} +  m^2 .
$
For this expression to be valid for ultrarelativistic particles there is an additional condition on the momenta: $|\vect k|/a(T) \ll (t_\text{int}^2 H)^{-1}$.

In principle, the perturbative evaluation of the propagators in curved spacetime can be performed in the same way as in Minkowski, taking into account the CTP doubling of the number of degrees of freedom. However, there are a couple of caveats. 

On the one hand, couple naive perturbation theory may be spoiled  both because of ultraviolet and infrared divergences.
While the leading ultraviolet divergences of the coupling parameters of an interacting field theory in a curved spacetime are the same as the divergences of the same field theory in flat spacetime,  there are subleading divergences which depend on the local curvature of the manifold. In general, in the spirit of effective field theories, one has to consider the most general Lagrangian which is compatible with the symmetries of the problem taking into account the gravitational background \cite{Bjerrum-BohrEtAl03, Donoghue94a,Donoghue94b}. Infrared divergences may also arise. Calculations which are formally correct might be spoiled by the divergent infrared behavior of the theory, even if final results are not explicitly infrared-divergent.  As opposed to ultraviolet divergences, which share a common structure for all theories, infrared divergences depend on the large scale structure of the spacetime, and therefore not many general things can be said about them. Infrared divergences usually appear when dealing with massless (or effectively massless) fields. Heuristically, one can make sure that the infrared divergences do not play an important role by checking that the relevant contribution to all intermediate expressions is not governed by the far infrared modes. If this is the case, one can be reasonably confident of the infrared stability of the results; otherwise one has to deal with the infrared divergences in a case by case basis.

On the other hand, in cosmological backgrounds even the computation of the free propagators is technically challenging. Exact closed analytic results can be obtained only in a few particular cases. The conformally coupled fields are one of this cases, in which the evaluation of the propagators reduces to the corresponding flat spacetime calculation through the introduction of the conformal time coordinate. (We will encounter a conformally coupled field in Sect.~IV.) Approximate closed analytic results can be also obtained under the adiabatic approximation. Whenever the energy of the mode is much larger than the expansion rate of the universe, the WKB solution provides a good representation for the propagator of the mode \cite{BirrellDavies}.  Thus, the adiabatic approximation, besides providing a well-defined particle interpretation, also facilitates the computation of the propagators.

\section{Interacting adiabatic particles in cosmology}

\index{Quasiparticle!adiabatic|(}

In Ref.~\cite{Arteaga08a} the propagation of quasiparticles in physical media from a second quantized perspective was analyzed using both a real-time approach and a frequency-based approach. In this section we generalize the real-time analysis to include adiabatic particles propagating in cosmological backgrounds. In detail, we investigate how the basic features characterizing the (quasi)particle propagation in cosmology can be extracted from the two-point correlation functions. We will concentrate on the novel aspects introduced by the universe expansion, skipping most technical details, which can be found in Ref.~\cite{Arteaga08a}.

 In cosmology scalar particles\footnote{For brevity, in the following we will not make the distinction between particles and quasiparticles, and will simply call ``particle'' every long-lived elementary excitation carrying momentum and energy, even if the background field is not necessarily in the adiabatic vacuum.} can be labeled by their conserved conformal momentum. When the energy of the particle is much higher than the expansion rate, adiabatic particles can be introduced. From a second quantized perspective, quasiparticles are characterized by their comoving momentum, their energy and their decay rate, this latter two quantities being time-dependent.
At this point we should make clear that with ``decay rate'' we refer to the rate at which the probability of finding the particle at a given comoving momentum decreases. Therefore having a non-vanishing decay rate does not necessarily mean that the particle decays into a lower mass state: ``decay'' can simply means that the particle has changed its momentum. The motivation for this terminology is the second quantized perspective, wich implies that we focus on the mode corresponding to the particle momentum rather than on the particle itself.

\subsection{Quantum states and energy of the excitations}

For a start, let us consider the quantum states corresponding to particle excitations in a cosmological context. Since by assumption the observation times are much longer than the interaction times, the construction of the quantum states will be based on the asymptotic in-out representation of the interacting fields, conveniently adapted to the expansion of the universe. Notice that here ``asymptotic'' refers to the properties of the interacting propagator, and not the cosmological model. Therefore we do not imply that the universe must have any asymptotic region: the asymptotic in-out representation of the fields is valid provided observation times are large enough, regardless of the details of the universe expansion.  

In the flat vacuum, the mode-decomposed interacting field operator, when acting on remote past or future times, and when evaluated inside a matrix element, can be approximated by a corresponding asymptotic free field operator \cite{WeinbergQFT,GreinerQFT}. In turn, this free field can be decomposed in creation and annihilation operators of the asymptotic particle states of physical momentum $\vect p$. Therefore the asymptotic representation for the field operator can be written
$
	\phi_\vect p \approx   [{Z}/(2 E_\vect p)]^{1/2} [ \hat a_\vect p+ \hat a^\dag_{-\vect p}],
$
where $E_\vect p$ is the physical energy of the particles  and  $Z$ is proportional 
to the  probability for the field operator to excite the vacuum with an energy $E_{\vect p}$. In Ref.~\cite{Arteaga08a} it is argued that an analogous construction can be extended to quasiparticle excitations in flat spacetime.
Here we wish to point out that the representation of the field operator in terms of creation and annihilation operators can be straightforwardly extended to cosmology as a function of the conformal momentum $\vect k$. The expression goes as follows:
\begin{equation}
	\phi_\vect k \approx \sqrt{\frac{Z_{\vect k}(t)}{2 E_\vect k(t)}} \big[ \hat a_\vect k(t) + \hat a^\dag_{-\vect k}(t)\big],
\end{equation}
where in this case $E_\vect k(t)$ is the physical energy of the particle excitation at time $t$ and  $Z_\vect k(t)$ is proportional\footnote{The precise definition goes as follows.  $Z_\vect k(t) \delta(E_{\beta}-E_{\vect k}(t)) = \sum_{\alpha,\beta} \rho_{\alpha\alpha}(t) |\langle \alpha|\hat\phi_{\vect k}|\beta\rangle|^2  \delta(E_{\beta}-E_{\alpha})$, where $|\alpha\rangle$ and $|\beta|\rangle$ are a complete set of orthornormal eigenvectors of the Hamiltonian spanning the Hilbert space. The density matrix $\hat\rho$ is assumed to be approximately diagonal in this basis. } 
to the  probability for the field operator to excite the background with an energy $E_{\vect k}(t)$. It must be stressed that the above representation is an asymptotic relation, valid only when evaluated inside a matrix element in the large time limit.  The parameter $Z_\vect k(t)$ can be  renormalized to one by rescaling the field; we will assume that such renormalization has been done in the following.

In a flat spacetime, the one-particle state is created by the action of the aysmptotic field operator on the vacuum: $|\vect k\rangle = \hat a^\dag_{\vect k}|0\rangle$. In Ref.~\cite{Arteaga08a} it was shown that when the vacuum is replaced by a slowly-varying background state $\hat\rho$ one can similarly find the one-particle state by acting with the creation operator on $\hat\rho$. The further generalization to cosmology is straightforward:
\begin{equation}
	\hat\rho^\pplus_\vect k(t) \approx \frac{1}{n_\vect k + 1}\, 
U(t,t_{0})\hat a^\dag_\vect k(t_{0}) \hat\rho(t_{0}) \hat a_\vect k(t_{0}) U(t_{0},t),
\end{equation}
where $n_\vect k$ is the occupation number of the mode with momentum $\vect k$ (an adiabatic invariant), $t_{0}$ is the initial creation time (assumed to be remote). The denominator ensures the proper normalization. The background state is assumed to evolve in timescales much longer than the interaction timescale.

The time evolution of the expectation value of the Hamiltonian is given by
\begin{equation}
	E^\pplus(t,t_{0};\vect k) =  \Tr{\big[\hat\rho^\pplus_\vect k(t) \hat H_\vect k(t)\big] } = \frac{1}{ n_\vect k + 1} \Tr{\big[ \hat\rho(t_{0})\hat a_\vect k(t_{0}) U(t_{0},t)\hat H_\vect k(t) U(t,t_{0})   \hat a_\vect k^\dag(t_{0}) \big] }.
\end{equation}
where $
	\hat H_\vect k(t) = E_{\vect k}(t) [\hat a_{\vect k}^\dag(t) \hat a_{\vect k}(t)+\hat a_{-\vect k}^\dag(t) \hat a_{-\vect k}(t)+1]
$ is the  Hamiltonian of the 2-mode $\pm \vect k$ (see Refs.~\cite{Arteaga07b,Arteaga08a} for a discussion of the convenience of including also the opposite momentum mode in general, although this is irrelevant in this presentation). Following similar steps as in Ref.~\cite{Arteaga08a}, we apply the Wick theorem, assuming that the background state $\hat\rho(t)$ is approximately Gaussian, and find:
\begin{equation*}
\begin{split}
	E^\pplus(t,t_{0};\vect k) &\approx \frac{1}{n_\vect k + 1} \Big\{
	 \Tr \big[ \hat \rho(t_{0}) \hat a_\vect k(t_{0}) \hat a_\vect k^\dag(t_{0}) 	\big] \Tr  \big[ \hat \rho(t_{0})U(t_{0},t) \hat H_\vect k(t) U(t,t_{0}) ]    \\
	&\quad+E_\vect k(t) \Tr  \big[ \hat \rho(t_{0}) \, \hat a_\vect k(t_{0}) U(t_0,t) \hat a_\vect k^\dag(t) U(t,t_0) \big] \\ &\qquad\times
	\Tr  \big[ \hat \rho(t_{0}) U(t_{0},t) \hat a_\vect k(t) U(t,t_{0}) \, \hat a_\vect k^\dag(t_{0}) \big] \Big\},
\end{split}
\end{equation*}
Expressing the creation and annihilation operators in terms of the asymptotic field and its derivative, we arrive at the following suggestive expression:
\begin{equation}\label{EnergyGuayAdiabatic}
	E^\pplus(t,t';\vect k) \approx E^{(0)}(t) +  \frac{E_\vect k(t)}{n_\vect k + 1}
	\left|\frac{[E_\vect k(t) + i\partial_t][E_\vect k(t') -i \partial_{t'}]}{2\sqrt{E_\vect k(t)E_\vect k(t')}} G_+(t,t';\vect k)\right|^2,
\end{equation}
where $E^{(0)}(t)=\Tr  \big[ \hat \rho(t_{0})U(t_{0},t) \hat H_\vect k(t) U(t,t_{0}) ]$ is the energy of the unexcited mode. Thus, the energy of the mode corresponding to the particle as a function of time can be computed from the positive Wightman function. Recall that for the above equation to be valid the condition $t_\text{obs} \gg t_\text{int}$ has to be fulfilled, where $t_\text{obs}$ the observation time is $t-t'$.

In the following we compute the evolution of the propagators [and hence the evolution of the expectation value of the energy, according to \Eqref{EnergyGuayAdiabatic}] for several timescale hierarchies.

\subsection{Short observation times}

Under the assumption of observation times much shorter than the Hubble timescale, $t_\text{obs} \ll H^{-1} $, one can find a time representation of 
the propagator by Fourier-transforming back Eq.~\eqref{shortime} . 
Since by hypothesis the observation times are also much larger than the interaction time, we can further approximate the 
self-energy by its value at the pole, similarly as in flat spacetime.
Assuming a small decay rate,  one finds \cite{ArteagaParentaniVerdaguer07,ArteagaThesis}
\begin{subequations}
\begin{align}\label{shortimetime}
	\bar G_\text{R}(t,t';\vect k) &= \frac{-i}{ E_\vect k(T)} \sin\left[E_\vect k(T)(t-t')\right] \expp{-\Gamma_\vect k(T)(t-t')/2} \theta(t-t'), \\
	\bar G^{(1)}(t,t';\vect k)    &= \frac{1+2n_\vect k}{ E_\vect k(T)} \cos\left[E_\vect k(T)(t-t')\right] \expp{-|\Gamma_\vect k(T)(t-t')/2|},
\end{align}
\end{subequations}
with $T=(t+t')/2$,
\begin{subequations}\label{RgammaShort}
\begin{equation} \label{R}
	E^2_\vect k(t) :=  m^2 + \frac{\vect k^2}{a^2(t)} + \Re\bar\SigmaR\boldsymbol(E_\vect k(t),t;\vect k\boldsymbol),
\end{equation}
and 
\begin{equation} \label{gamma}
	\Gamma_\vect k(t) := -\frac{1}{E_\vect k(t)} \Im\bar\SigmaR\boldsymbol(E_\vect k(t),t;\vect k\boldsymbol).
\end{equation}
\end{subequations}
Introducing these expressions in \Eqref{EnergyGuayAdiabatic} and taking into account that $\bar G_+(t,t') = [\bar G^{(1)}(t,t')+\bar \GR(t,t')]/2$ for $t>t'$, we obtain:
\begin{equation}
	E(t,t';\vect k) = E^{(0)}(t) + E_\vect k(t) (1+n_\vect k) \expp{-\Gamma(T)(t-t')}.
\end{equation}
Therefore the energy of the quasiparticle excitation at time $t$ is $E_\vect k(t)$, and it decays in a timescale $\Gamma(T)$. The factor  $1+n_\vect k$, which is relevant only for non-vacuum states, is a consequence of the fact that slightly more than one quasiparticle is excited because of the Bose-Einstein statistics (see Ref.~\cite{Arteaga08a}).

The short-time approximation  is valid provided the following scale separation is verified:
$
	E_\vect k^{-1}, t_\text{int} \ll t_\text{obs} \ll H^{-1}.
$
A careful analysis \cite{ArteagaParentaniVerdaguer07,ArteagaThesis} shows  that for ultrarelativistic particles there is an additional condition on the momenta: $|\vect k|/a(T) \ll (t_\text{int}^2 H)^{-1}$.

\subsection{Long observation times}

When the observation times are of the order of the expansion timescale we can no longer use Fourier-transform methods. Instead, we will look for a suitable WKB approximation for the propagators.
Let us start by considering the equation of motion of the interacting propagator.
By acting with the differential operator 
\[
	\frac{1}{a^3(t)} \derp{}{t}{} \left(a^3(t) \derp{}{t}{} \right) + m^2 + \xi R(t) + \frac{\vect k^2}{a^2(t)}
\]
(with $R(t)$ being the Ricci scalar and $\xi$ being the conformal coupling parameter) on equation \eqref{Dyson}, we get the equation of motion for the retarded propagator:
\begin{equation} \label{EqRet}
\begin{split}
	&\left[ \frac{1}{a^3(t)} \derp{}{t}{} \left(a^3(t) \derp{}{t}{} \right) + m^2 + \xi R(t) + \frac{\vect k^2}{a^2(t)} \right]
	\left[ \frac{\bar\GR(t,t';\vect k)}{a^{3/2}(t) a^{3/2}(t')} \right]  \\
	&\qquad\qquad + \frac{1}{{a^{3/2}(t) a^{3/2}(t')} } \int \ud{s} \bar\SigmaR(t,s;\vect k)  \bar\GR(s,t';\vect k) = \frac{-i}{a^3(t)} \delta(t-t')
\end{split}
\end{equation}
This equation so far is exact. 
Since we are interested in a first order adiabatic solution, 
we start by discarding all terms of the equation of motion which are of higher 
order in $H/E_\vect k$:
\begin{equation} \label{eqc}
\begin{split}
	&\left[ \derp[2]{}{t} + m^2 + \frac{\vect k^2}{a^2(t)} \right]
	 \bar\GR(t,t';\vect k)  + \int_{t_{0}}^t \ud{s} \bar\SigmaR(t,s;\vect k)  \bar\GR(s,t';\vect k) = -i \delta(t-t').
\end{split}
\end{equation}
The equation of motion for the retarded propagator in the rescaled time $u = Ht$ is:
\begin{equation*}
\begin{split}
	\Big[ H^2 \derp[2]{}{u} &+ m^2 + \frac{\vect k^2}{a^2(u)} \Big]
	 \bar\GR(u,u';\vect k) + \frac{1}{H} \int_{u_{0}}^u \ud{v} \bar\SigmaR(u,v;\vect k)  \bar\GR(v,u';\vect k) = -i H \delta(u-u').
\end{split}
\end{equation*}
Expanding the non-local term as 
\begin{equation} \label{expansion}
\begin{split}
	N(u,u') &:=\frac{1}{H} \int_{u_{0}}^u  \ud{v} \bar\SigmaR(u,v;\vect k)  \bar\GR(v,u';\vect k)\\ &=: \left[ \delta E^2_\vect k(u) + H \Gamma_\vect k(u) \derp{}{u} \right] \bar\GR(u,u';\vect k) + O(H^2),
\end{split}
\end{equation}
the equation of motion  can be approximated by
\begin{equation} \label{operatorRetardedAdiabatic}
	\left[ H^2 \derp[2]{}{u} + H \Gamma_\vect k(u) \derp{}{u} + E_\vect k^2(u)\right]
	 \bar\GR(u,u';\vect k)  = -i H \delta(u-u'),
\end{equation}
where $E_\vect k^2(u) = m^2 + \vect k^2 / a^2(u) + \delta E^2_\vect k(u)$. The leading order adiabatic solution is (assuming that $\Gamma_\vect k$ is much smaller than $E_\vect k$)
\begin{equation*}
\begin{split}
	\bar G_\text{R}(u,u';\vect k) &= \frac{-i}{  \sqrt{ E_\vect k(u) E_\vect k(u')}} \sin\left({\frac{1}{H}\int^{u}_{u'} \ud{v} {E_\vect k(v)}} \right) \expp{-{\frac{1}{2H}\int^{u}_{u'} \ud{v} \Gamma_\vect k(v)} }  \theta(u-u'),
\end{split}
\end{equation*}
or, going back to the original time $t$,
\begin{equation}\label{InteractingWKB}
\begin{split}
	\bar G_\text{R}(t,t';\vect k) &= \frac{-i}{  \sqrt{ E_\vect k(t) E_\vect k(t')}} \sin\left({\int^{t}_{t'} \ud{s} {E_\vect k(s)}} \right) \expp{-{{}\int^{t}_{t'} \ud{t} \Gamma_\vect k(t)/2} }  \theta(t-t').
\end{split}
\end{equation}
The interacting propagator can be expressed in terms of the adiabatic evolution of the quasiparticle energy and decay rates. Similarly, the Hadamard propagator can be expressed as \cite{ArteagaThesis}:
\begin{equation}\label{HadamardAdiabatic}
\begin{split}
	\bar G^{(1)}(t,t';\vect k) = \frac{1+2n_\vect k}{  \sqrt{ E_\vect k(t) E_\vect k(t')}} \cos\left({\int^{t}_{t'} \ud{s} E_\vect k(s)}\right) \expp{-{{}|\int^{t}_{t'} \ud{t} \Gamma_\vect k(t)/2|} }.
\end{split}
\end{equation}
\index{Decay rate}

Taking into account that $G_+(t,t') = [\GR(t,t') + G^{(1)}(t,t')]/2$ for $t>t'$, and introducing Eqs.~\eqref{InteractingWKB} and \eqref{HadamardAdiabatic} in \eqref{EnergyGuayAdiabatic} we find in the leading adiabatic order:
\begin{equation} 
	E^\pplus(t,t_{0};\vect k) \approx E^{(0)} (t)+ E_\vect k(t)(1+n_\vect k)\expp{-\int_{t'}^t \ud s \Gamma(s)}.
\end{equation}
From this equation we identify $E_\vect k(t)$ as the energy of the quasiparticles, which is slowly evolving with the background, and $\Gamma_\vect k(t)$ as their net decay rate. Recall that the factor $(1+n_\vect k)$ is a consequence of the fact that the quasiparticle state actually contains more than one particle excitation. Recall also that in this context decay does not necessarily mean the particle going into a lower mass state, but it means that the particle changes its momentum.

Let us see now how the energy and the decay rates can be extracted from the propagators, by considering separately two different situations.

\paragraph{Short interaction times: $E_\vect k^{-1}, t_\mathrm{int} \ll t_\mathrm{obs} , H^{-1}$.} When the interaction times are much shorter than the inverse expansion rate the particle energy and the decay rate can be  computed from Eqs.~\eqref{R} and \eqref{gamma}, respectively \cite{ArteagaParentaniVerdaguer07,ArteagaThesis}. In other words, as far as the interaction process is considered the universe expansion can be neglected. 

For non-relativistic particles the requirement that the interaction time is small as compared to the typical expansion time can be somewhat relaxed, since the precise condition for Eqs.~\eqref{R} and \eqref{gamma} to be valid is:
\begin{equation}
	\frac{t_\text{int}}{E_\vect k}\derp{E_\vect k}{t} \sim \frac{\vect k^2/a^2}{E^2_\vect k} t_\text{int} H \ll 1.
\end{equation}

\paragraph{Long interaction times: $E_\vect k^{-1} \ll t_\mathrm{int},  H^{-1} \ll t_\mathrm{obs}$.}
When the interaction times  are of the 
 order of the expansion timescale $H^{-1}$, the expansion effects cannot be neglected during the interaction time, and the Fourier transform is no longer useful. To investigate how $\delta E_\vect k$ and $\Gamma_\vect k$ can be expressed in terms of the self-energy, we reconsider the non-local term:
\begin{equation*} 
\begin{split}
	N(u,u')&= \frac1H \int_{u_{0}}^u  \ud{v}\frac{-i \bar\SigmaR(u,v;\vect k)}{  \sqrt{ E_\vect k(v) E_\vect k(u')}} \sin\left({\frac{1}{H}\int^{v}_{u'} \ud{v'} {E_\vect k(v')}} \right) \expp{-{\int^{v}_{u'} \ud{v'} \Gamma_\vect k(v')/(2H)} }  \theta(v-u').
\end{split}
\end{equation*}
Splitting the interval $[u',v]$ in  $[u,v]$ minus $[u,u']$ yields
\begin{equation*} 
\begin{split}
	N(u,u')&= \frac1H \int_{u'}^u  \ud{v}\frac{-i \bar\SigmaR(u,v;\vect k)}{  \sqrt{ E\ret_\vect k(v) E\ret_\vect k(u')}}\expp{-{\int^{v}_{u'} \ud{v'} \Gamma_\vect k(v')/(2H)} }  \\
	 &\qquad \bigg[ \sin\left({\frac{1}{H}\int^{u}_{u'} \ud{v'} {E\ret_\vect k(v')}} \right)  \cos\left({\frac{1}{H}\int^{u}_{v} \ud{v} {E\ret_\vect k(v')}} \right) \\ &\qquad- \cos \left({\frac{1}{H}\int^{u}_{u'} \ud{v'} {E\ret_\vect k(v')}} \right) \sin\left({\frac{1}{H}\int^{u}_{v} \ud{v'} {E\ret_\vect k(v')}}  \right) \bigg].
\end{split}
\end{equation*}
Replacing $v$ by $u$ in the argument of the exponential function (the total amount of decay during the interaction process is negligible; otherwise the particle would have completely decayed at the observation point), we find:
\begin{equation}  \label{NonLocalGeneralL}
\begin{split} 
	N(u,u')&\approx (\delta E\ret_\vect k)^2(u) \bar\GR(u,u';\vect k) +H \Gamma\ret_\vect k(u)  \derp{\bar\GR(u,u';\vect k)}{u},
\end{split}
\end{equation}
where in non-rescaled time $t$,
\begin{subequations}\label{GammaRSuperGen}
\begin{align}
	(\delta E\ret_\vect k)^2(t) &= \int_{t'}^t \ud s   \frac{E_\vect k\ret(t)\bar\SigmaR(t,s;\vect k) }{\sqrt{E\ret_\vect k(t)E\ret_\vect k(s)}} \cos\left({\int^{t}_{s} \ud{s'} {E\ret_\vect k(s')}} \right), \\
	\Gamma\ret_\vect k(t) &= -\int_{t'}^t \ud {s} \frac{\bar\SigmaR(t,s;\vect k)}{\sqrt{E\ret_\vect k(t)E\ret_\vect k(s)}}   \sin\left({\int^{t}_{s} \ud{s'} {E\ret_\vect k(s')}} \right).\label{GammaSuperGen}
\end{align}
\end{subequations}
Eqs.~\eqref{GammaRSuperGen} give the most general representation of the energy shift and the decay rate in terms of the self-energy.  Notice that the self-energy is evaluated in a kind of frequency representation, but with the  frequency varying along the interaction range, because the on-shell position changes significantly during the interaction process. Notice also the presence of the square root prefactor, which can be interpreted as the geometric mean of the particle redshift along the interaction time.

Let us concentrate on the dissipative effects in the remaining of this section. When the interaction times are comparable to the inverse expansion rate, the interpretation of \Eqref{GammaSuperGen} as a decay rate may be hindered by the universe expansion. In effect, since that the decay rate is only a meaningful concept when considering observation times much larger than the interaction time, the decay rate is also expected to evolve on scales much larger than the typical expansion time. However, the decay rate extracted from \Eqref{GammaSuperGen} evolves on timescales comparable to the universe expansion because of the redshifting. To avoid these difficulties, let us consider the decay rate in the particle rest frame, which naturally can be conjectured to be:
\begin{equation}
		\gamma_\vect k(t) = -\frac1m\int_{t'}^t \ud {s} {\bar\SigmaR(t,s;\vect k)}  \sin\left({\int^{t}_{s} \ud{s'} {E\ret_\vect k(s')}} \right).
\end{equation}
 Recalling that this quantity is only meaningful when observed for large periods of time, let us average the decay rate for timescales much smaller than the observation time. In other words, we choose a timescale $t_\text{av}$ much larger than $t_\text{int}$ but still much smaller than $t_\text{obs}$, and average the decay rate over that scale:\begin{equation}
\begin{split}
	\gamma_\vect k(t)& \approx \frac{1}{t_\text{av}}\int_{t-t_\text{av}/2}^{t+t_\text{av}/2} \ud{t''} \gamma\ret_\vect k(t'') \\ &\approx-\frac{1}{m\, t_\text{av}}\int_{t-t_\text{av}/2}^{t+t_\text{av}/2}\ud{t''}  \int_{t'}^{t''} \ud {s} {\bar\SigmaR(t'',s;\vect k)}   \sin\left({\int^{t''}_{s} \ud{s'} {E\ret_\vect k(s')}} \right).
\end{split}
\end{equation}
Taking into account that $t_\text{av}$ is much larger than the interaction time, and introducing the semisum and semidifference coordinates we get:
\begin{equation*}
\begin{split}
	\gamma_\vect k(t) 
	  &\approx-\frac{1}{m\, t_\text{av}}\int_{t-t_\text{av}/2}^{t+t_\text{av}/2}  \ud{T}\int \ud {\Delta} {\bar\SigmaR(T+\Delta/2,T-\Delta/2;\vect k)} \sin\left({\int^{T+\Delta/2}_{T-\Delta/2} \ud{s'} {E\ret_\vect k(s')}} \right),
	  \end{split}
\end{equation*}
or, taking into account that $\SigmaR$ does not depend significantly on $T$ during the averaging time,
\begin{equation}
\begin{split}
	\gamma_\vect k(t) 
	  &\approx-\frac1m\Im \int \ud {\Delta}  {\bar\SigmaR(t+\Delta/2,t-\Delta/2;\vect k)} \expp{i\int^{t+\Delta/2}_{t-\Delta/2} \ud{s'} {E_\vect k(s')}}.
	  \end{split}
\end{equation}
In terms of the following ``improved'' frequency representation for the self-energy,
\begin{equation}\label{frequencyImproved}
	\SigmaImp([E_\vect k],t;\vect k) := \int \ud \Delta \bar\SigmaR(t+\Delta/2,t-\Delta/2;\vect k) \expp{i\int_{t-\Delta/2}^{t+\Delta/2} \ud s E_\vect k(s)}
\end{equation}
the decay rate in the particle rest frame can be expressed as
\begin{equation} \label{gammaImproved}
	\gamma_\vect k(t) \approx -\frac1m \Im\SigmaImp([E_\vect k],t;\vect k). \end{equation}
Notice that the improved frequency representation has been defined only for on-shell values of the frequency; in this sense it cannot be considered a proper integral transform. Note also that  the improved frequency representation yields the same results as the standard frequency representation for comoving particles.

\index{Quasiparticle!adiabatic|)}
\index{Quasiparticle!in curved spacetime|)}
\index{Propagator!adiabatic|)}

\section{Expansion-induced decay in cosmology}

Let us apply the techniques shown in  the previous section to analyze  in a particular model the dissipative effects on the propagation in an expanding universe.
Let us consider two massive fields with large masses but with a small mass difference $\dm := M-m \ll M $. 
As it is shown in Refs.~\cite{Parentani95,ArteagaParentaniVerdaguer05}, the model can be 
interpreted as a 
field-theory description of a 
relativistic
two-level atom 
(of rest mass $m$ and energy gap $\dm$)
interacting with a 
scalar massless radiation field. 
The masses of the fields will be assumed to be much larger than the expansion rate of the universe, allowing us to introduce the adiabatic approximation. The mass gap between the two massive states will be taken to be of the order of the expansion rate of the universe.

In detail, the model consists of
two massive fields $\phim$, 
and $\phiM$, interacting with a massless field, $\chi$, via a trilinear coupling. 
The total action  can be decomposed as $S = S_m + S_M + S_\chi + S_\text{int}$, where:
\begin{subequations}
\begin{align}
	S_m &= \frac{1}{2} \int \ud t \ud[3] {\vect x} a^3(t)  \left( (\partial_t \phim)^2 - \frac1{a^2(t)} (\partial_\vect x \phim)^2 - m^2 \phim^2 \right),\\
	S_M &= \frac{1}{2} \int \ud t \ud[3] {\vect x} a^3(t)  \left( (\partial_t \phiM)^2 - \frac1{a^2(t)} (\partial_\vect x \phiM)^2 -  M^2 \phiM^2 \right),\\
	S_\chi &= \frac{1}{2} \int \ud t \ud[3]{\vect x} a^3(t) 
	\left(  (\partial_t \chi)^2 - \frac1{a^2(t)} 
	(\partial_\vect x \chi)^2 - \xi R(t) 
	\chi^2 \right) ,\\
	S_\text{int} &= g M \int \ud t \ud[3]{\vect x} a^3(t)  \phim \phiM \chi,
\end{align}
\end{subequations}
with $R(t)$ being the Ricci scalar. We 
assume that the massless field is conformally coupled to gravity, so that $\xi = 1/6$.
In this paper we will be interested in studying the vacuum effects, so that the massless field will be assumed to be in the conformal vacuum and the massive fields in the adiabatic vacuum.; see Refs.~\cite{ArteagaParentaniVerdaguer05,ArteagaParentaniVerdaguer07} for an analysis of the temperature effects.

\begin{figure}
	\centering
	\includegraphics[width=0.40\textwidth]{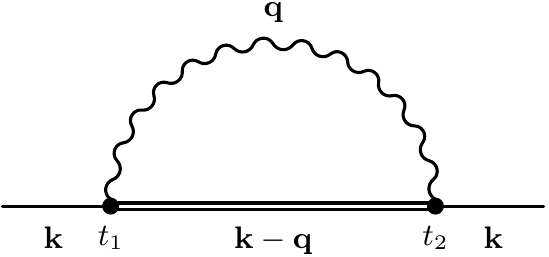}
	\caption{Feynman diagram leading to the one-loop self-energy of the less massive particle. Straight lines represent the fundamental sate, double lines the excited state and curly lines the massless particle.} \label{fig:SigmaCurved}
\end{figure}

The goal is to extract the dissipative effects on the propagation of the unexcited atom from the self-energy. To one loop, the self-energy can be computed by using CTP perturbation theory as (see figure \ref{fig:SigmaCurved})
\begin{equation}
	- i \bar\Sigma^{ab}(t_1,t_2;\vect k) ={(ig M)^2} c^{aa} c^{bb} \int \udpi[3]{\vect q} \bar G^{*(0)}_{ab}(t_1,t_2;\vect k-\vect q)  \Delta^{(0)}_{ab}(t_1,t_2;\vect q) 
\end{equation}
(no summation implied), where $G^{*(0)}_{ab}(t_1,t_2;\vect k-\vect q)$ are the free adiabatic propagators of the more massive field and $\Delta^{(0)}_{ab}(t_1,t_2;\vect q)$ are the free conformal propagators of the massless field \cite{BirrellDavies,ArteagaParentaniVerdaguer07}. We recall that $a,b,c\ldots$ indices refer to the different CTP branches and that $c^{ab}= \text{diag}(1,-1)$. The retarded self-energy corresponds to $\SigmaR(t_1,t_2;\vect k) = \Sigma^{11}(t_1,t_2;\vect k) - \Sigma^{12}(t_1,t_2;\vect k)$. 
Since we are investigating the effects induced by the universe expansion, and therefore the interaction time is of the order of the inverse Hubble rate, we will  evaluate the self-energy in the improved frequency representation.

\begin{figure}
	\centering
  \includegraphics[width=0.25\textwidth]{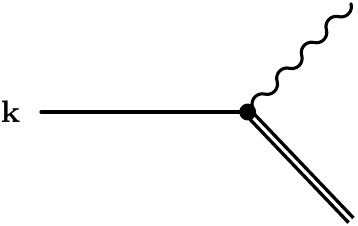}
	\caption{Feynman diagram contributing to the decay rate of the field $\phim$. The diagram violates energy conservation, and it is hence forbidden in flat spacetime. However this needs not be the case in an expanding universe.}\label{fig:ImSigmaCosCurved}
\end{figure}

The imaginary part
\footnote{Notice that here we are abusing the notation because in the time representation the retarded self-energy is purely real. ``Imaginary part'' here refers to the standard or improved frequency representations. In the time representation, it corresponds to the odd part of $\bar\SigmaR$ under the exchange of $t_1$ and $t_2$.}  
of the self-energy is given by 
\begin{equation}
	\Im\bar\SigmaR(t_1,t_2) = \frac{1}{2i} \left[ \bar\Sigma^{12}(t_1,t_2) -\bar\Sigma^{21}(t_1,t_2) \right] \, .
\end{equation}
By applying the CTP Feynman rules in the physical time 
representation to the diagram in figure \ref{fig:SigmaCurved}, 
to order $g^2$ we get:
\begin{equation} \label{ImSigmaCos1}
\begin{split}
	\bar\Sigma^{21} (t_1,t_2;\vect k) &= 
	\frac{i g^2  M^2 
	}{ a(t_1)a(t_2)} \int \udpi[3] {\vect q} \frac{\expp{ -i \int_{t_2}^{t_1} \ud {t'} E_\vect{k-q}^*(t') -i\int_{t_2}^{t_1} \ud t' |\vect q|/a(t')}}{ 2\sqrt{ E_\vect{k-q}^* (t_1)E_\vect{k-q}^*(t_2)}} \frac{1}{2|\vect q|}  \end{split}
\end{equation}
with $E^{*2}_{\vect k-\vect q}(t)=(\vect k-\vect q)^2/a^2(t) +  M^2$. 
\Eqref{ImSigmaCos1} can be interpreted in terms of the square of the Feynman diagram represented in figure \ref{fig:ImSigmaCosCurved},
corresponding to the emission of a photon (this 
diagram would  
vanish on shell in flat spacetime). In the (standard or improved) frequency representations,
\begin{equation}
\begin{split}
	\widetilde\Sigma^{21} ([E_\vect k],T;\vect k) \approx- 2i \Im \SigmaImp ([E_\vect k],T;\vect k).
\end{split}
\end{equation}
The equality is valid because the Fourier transform of $\bar\Sigma^{12}$ gives exponentially damped contributions  in the high mass limit we are using. Therefore
\begin{equation} \label{ImSigmaCos}
\begin{split}
	\Im \SigmaImp ([E_\vect k],T;\vect k) &= -g^2  M^2 
	 \int \frac{\ud \Delta }{ 2a(t_1)a(t_2)}\int \udpi[3] {\vect q} \frac{\expp{ i \int_{t_2}^{t_1} \ud {t'}[E_\vect{p}(t')- E_\vect{k-q}^*(t')- |\vect q|/a(t')]}}{ 2\sqrt{ E_\vect{k-q}^* (t_1)E_\vect{k-q}^*(t_2)}} \frac{1}{2|\vect q|} .\end{split}
\end{equation}
where $t_1 = T+\Delta/2$ and $t_2 = T-\Delta/2$. Noticing that it vanishes in the case of flat spacetime, let us evaluate this quantity for two different models of the universe expansion.

\subsection{Vacuum effects in de Sitter}

Let us consider flat coordinates in the de Sitter spacetime, which corresponds to a flat cosmological model with 
\begin{equation}
	a(t) = a(T) \expp{H(t-T)}.
\end{equation}
For the sake of simplicity, we will show the details of the calculation in the comoving case, and simply quote the result for the case of arbitrary geodesic motion. 

From \Eqref{ImSigmaCos}, the vacuum contribution to the imaginary part of the self-energy in de Sitter is given by:
\begin{equation}
\begin{split}
	\Im\bar\SigmaR(m,T;\vect 0) &= -\int \ud \Delta \int_{0}^\infty \ud k \frac{g^2 m k}{16 \pi^2  a^2(T)} \\ &\qquad\times \exp{\left[-i\int_{T-\Delta/2}^{T+\Delta/2}\ud t\left( \dm + \frac{k}{a(T)} \expp{H(T-t)} \right)\right] }.
\end{split}
\end{equation}
Integrating the argument of the exponential yields,
\begin{equation*}
	\Im\bar\SigmaR(m,T;\vect 0) = -\int \ud \Delta \int_{0}^\infty \ud k \frac{g^2 m k}{16 \pi^2  a^2(T)} \expp{-i \dm \Delta - \frac{ik}{a(T)H} \sinh\left( \frac{H\Delta}{2} \right)  - \epsilon k },
\end{equation*}
where have added a small $-\epsilon k$ term to ensure convergence of the $k$ integral: 
\begin{equation*}
	\Im\bar\SigmaR(m,T;\vect 0) = \int \ud \Delta 
	\frac{g^2 m H^2}{16\pi^2} \frac{\expp{(-i \dm + H) \Delta}}{(-1+\expp{H\Delta}-i \epsilon)^2}.
\end{equation*}
The $\Delta$ integration can calculated by residues in the complex plane, closing the circuit through a parallel line passing by $2\pi i /H$ and taking into account the pole at $\Delta=i\epsilon$. The result is the following:
\begin{equation}
	\Im\bar\SigmaR(m,T;\vect 0) = - \frac{g^2}{8\pi}  \frac{m \dm}{\expp{2\pi\dm /H}-1} = - \frac{g^2}{8\pi}  m \dm n_{H/2\pi}(\dm),
\end{equation}
where $n_T(x)$ is the Bose-Einstein function corresponding to a physical temperature $T$. This results coincides with the imaginary part of the self-energy in a Minkowski thermal bath at physical temperature $H/(2\pi)$ \cite{ArteagaParentaniVerdaguer05,ArteagaParentaniVerdaguer07}. A decay rate can be associated to this imaginary part of the self-energy:
\begin{equation}
	\Gamma_\vect 0(T) = \gamma_\vect 0(T)  = - \frac{1}{m} \Im\bar\SigmaR(m,T;\vect 0) 	 = \frac{g^2}{8\pi}   \dm n_{H/2\pi}(\dm).
\end{equation}
Recall that in the comoving case there is no difference between the standard and improved self-energies.
\index{Decay rate}

\index{de Sitter effective temperature}
The result is not at all unexpected, since it is well known \cite{BirrellDavies} that a comoving particle detector in de Sitter conformal vacuum perceives a bath of thermal radiation at a temperature  $H/(2\pi)$. The fact that the decay rate has an exact thermal character can be traced to the existence of thermalization theorems under the presence of event horizons \cite{Takagi86}. Our results are also consistent with those of Bros \emph{et al.}~\cite{BrosEtAl07}, who similarly find inestabilities for massive particles propagating in de Sitter, using a global rather than an adiabatic approach to the particle concept.

This result can be extended to moving particles with arbitrary momentum. Notice that in this case it is important to use the improved frequency representation because the interaction timescale is of order $H^{-1}$.  We have evaluated the integrals in a non-relativistic expansion in $\vect k^2/m^2$. Interestingly, we have found that the additional momentum-dependent terms cancel and that the result is 
\begin{equation}
	\Im\SigmaImp([E_\vect k],T;\vect k) = \frac{g^2}{8\pi}  m \dm\, n_{H/2\pi}(\dm)
\end{equation}
at least to order $\vect k^{10}/m^{10}$. It is therefore reasonable to assume that the above result is valid to all orders in perturbation theory. The result is not unexpected, since the decay rate in the particle rest frame 
\begin{equation}\label{decayRateDeSitter}
	\gamma = -\frac{1}{m} \Im\SigmaImp([E_\vect k],T;\vect k)  = \frac{g^2}{8\pi}  \dm\, n_{H/2\pi}(\dm)
\end{equation} 
is time- and momentum- independent as required by the de Sitter invariance of the problem (bear in mind that both the adiabatic and conformal vacua are de Sitter invariant). Since the invariance is not manifest from our expressions, this provides a non-trivial check of the calculation. We have also computed the decay rate in the comoving frame, according to \Eqref{GammaRSuperGen}. The result is the following:
\begin{equation}
	\Gamma_\vect k(t) = \frac{g^2}{8\pi}  \dm\, n_{H/2\pi}(\dm) \left( 1 - \frac{\vect k^2}{2 a^2(t) m^2} \right) + O(\vect k^4/m^4).
\end{equation}

\subsection{Vacuum effects in power-law inflation}

Let us now consider the case of power-law inflation, where the scale factor evolves according to
\begin{equation}
	a(t) = a(T) \left( \frac{t}{T} \right)^\alpha.
\end{equation}
The Hubble rate is given by $H(t) =\dot a(t) /a(t) = \alpha/t$. Since expressions rapidly become  cumbersome, we shall only sketch the calculation in the simplest situations, and will present the result for the other cases.

As before, let us first consider comoving particles. According to Eq.~\eqref{ImSigmaCos}, the imaginary part of the self-energy is given by
\begin{equation}
\begin{split}
	\Im\bar\SigmaR(m,T;\vect 0) &= -\int \ud \Delta \int'^\infty \ud k \frac{g^2 m k}{16 \pi^2  a^2(T)} \left( 1- \frac{\Delta^2}{4T^2} \right)^{-\alpha} \\ &\qquad\times \exp{\left\{-i\int_{T-\Delta/2}^{T+\Delta/2} \ud t \left[ \dm + \frac{k}{a(T)} \left( \frac{T}{t} \right)^\alpha \right]\right\} }.
\end{split}
\end{equation}
Integrating the argument of the exponential, and  performing the integral over $k$ yields (it is necessary to add a  $-\epsilon k$ term to ensure convergence)
\begin{equation}
\begin{split}
	\Im\bar\SigmaR(m,T;\vect 0) &= \frac{2 g^2 m (-1+\alpha)^2}{\pi^2} \int \ud \Delta 
	\expp{-i \dm \Delta} \\
	&\qquad \times\frac{ 
	\left( 1- \frac{\Delta^2}{4T^2} \right)^{-\alpha}
	(2T + \Delta)^{2\alpha}}
	{\{ T^\alpha(2 T + \Delta) - (2T-\Delta)^{1-\alpha}
	[T(2T+\Delta)]^\alpha\}^2}.\raisetag{4em}	
\end{split}
\end{equation}
In order to proceed further a particular value of $\alpha$ has to be chosen. Closed analytic expressions can be obtained for positive integers values of $\alpha$. For the case of $\alpha=4$ we obtain:
\begin{equation*}
\begin{split}
	\Im\bar\SigmaR(m,T;\vect 0) &= \frac{9g^2 m }{16\pi^2} \int \ud \Delta 
	\frac{ \expp{-i \dm \Delta} 
	\left(-4T^2 + \Delta^2\right)^4}
	{\Delta ^2 \left(2 T^2+\Delta ^2\right)^2}, \quad \alpha=4.
\end{split}
\end{equation*}
This function can be integrated in the complex plane by considering the residues of the double pole located at $\Delta = -2\sqrt{3} i T$.\footnote{The other double pole would be located at $\Delta = i\epsilon$ had we kept track of the $\epsilon$ terms, and therefore does not have to be taken into account.} The result is the following:
\begin{equation}\label{decayRatePowerLaw}
	\Im\bar\SigmaR(m,T;\vect 0) = -\frac{g^2 }{2 \pi } m  \dm \expp{-2 \sqrt{3} \dm T} =-\frac{ g^2 }{2 \pi }m \dm \expp{-{8 \sqrt{3} \dm}/{H(T)}}, \quad \alpha = 4.
\end{equation}

This expression can be generalized, on the one hand, by considering different values of $\alpha$, and on the other hand by considering particles in motion.  We have found that at this order there is only a non-vanishing contribution to the imaginary part of the self-energy for even values of $\alpha$ larger than two. In the case $\alpha = 6$ the following result is found:
\begin{equation}
	\Im\bar\SigmaR(m,T;\vect 0) =-\frac{\left(3+\sqrt{5}\right)g^2 }{4 \pi }m \dm \expp{-12 \sqrt{5+2 \sqrt{5}}
   \dm T} , \quad \alpha = 6.
\end{equation}
Expressions become increasingly cumbersome with $\alpha$. The momentum corrections can be found in Ref.~\cite{ArteagaThesis}.

The decay rates can be immediately extracted from the imaginary part of the self-energy following Eq.~\eqref{gammaImproved}. Qualitatively, the behavior is the same as in de Sitter: the vacuum-induced decay rate is exponentially suppressed except when $H \gtrsim \dm$. However, in this case the rates do not have an exact thermal character and depend on time, because the Hubble rate varies during the universe evolution.
Unfortunately we have not been able to compute the decay rate in the ultrarelativistic limit, which is the most interesting limit (because even slowly moving particles will become rapidly moving with respect the comoving frame as we go back in time along the universe expansion).

\section{Summary and discussion}

In this paper we have analyzed the adiabatic particle excitations of an interacting field in a cosmological background. In particular, we have shown that the time evolution of the energy of interacting adiabatic particles can be expressed in terms of the two-point correlation functions [Eq.~\eqref{EnergyGuayAdiabatic}]. These in turn can be calculated in a WKB approximation---see Eqs.~\eqref{InteractingWKB}. The long-time dynamics of the particle can be essentially characterized from a second quantized perspective by two semi-local quantities, the energy shift and the net decay rate, this latter quantity being understood as the total transition rate to any other field state. 

Two different physical situations have been distinguished. If the interaction timescale is much smaller than the typical expansion rate, the energy shift and the decay rate can be extracted from the retarded self-energy in a quasilocal frequency representation [Eqs.~\eqref{RgammaShort}]. If, on the contrary, the interaction timescale is comparable to the expansion rate, then the energy shift and the decay rate can be extracted from a kind of frequency transform which takes into account the evolution of the on-shell condition during the interaction time [\Eqref{GammaRSuperGen}]. We also defined an improved frequency representation, \Eqref{frequencyImproved}, from which the decay rate in the particle rest frame can be extracted. 

These results have been applied to the analysis of the evolution of a doublet of massive fields interacting with a third massless field. We have focused on the dissipative effects generated by the universe expansion, which are relevant when the expansion rate of the universe is of the order of the mass gap or larger. In the case of de Sitter,  the decay rate corresponds to that of a particle at rest in a flat thermal bath at the de Sitter effective temperature [\Eqref{decayRateDeSitter}]; moreover we verified that the decay rate is de Sitter invariant. In the case of power-law inflation, a qualitatively similar decay rate is found [Eq.~\eqref{decayRatePowerLaw}].

The approach presented in this paper is complementary to other analysis of the cosmological decay rates in the literature \cite{Tsaregorodtsev95,AudretschSpangehl85,AudretschEtAl87,BrosEtAl07}, which have been based on asymptotic or symmetry-based particle concepts and S-matrix computations. Concerning the operational definition of the particle concept in terms of the response of a particle detector, in Ref.~\cite{ArteagaThesis} it is argued that  the operational and adiabatic approaches lead to identical results within the range of validity of the adiabatic approximation, even if self-interaction is taken into account.

A key point in this paper has been using several approximations controlled by different expansion parameters. The adiabatic approximation, controlled by the ratio $H/E_\vect k$, provided us with a well defined particle concept and furthermore allowed having simple analytic expressions. The asymptotic approximation, controlled by the ratio $t_\text{int}/t_\text{obs}$, allows to obtain model-independent predictions on the evolution of the propagators. The Gaussian approximation, which is formally controlled by $1/N$ (with $N$ being the number of fields), reduces the problem of computing the evolution of the quantum state of the particle to the computation of the two-point correlation functions.  Most of the times we have not explicitly indicated the expansion parameter of the different approximations for the sake of brevity. 

One of the requirements of the calculation of the dissipative effects in the adiabatic vacuum is that observation times must be much larger than interaction times. Since interaction times are of the order of the Hubble time $H^{-1}$, interesting results can only be obtained when the Hubble rate is slowly varying, or, in other words, when $\dot H \ll H^2$. This condition can only be achieved in inflationary contexts. Notice that in the power-law inflationary model $\dot H = H^2/\alpha$, so that large values of $\alpha$ should be considered (although for computational simplicity we just studied the $\alpha=4,6$ cases). Concerning also this inflationary model, recall that no contribution to the decay rate was found for odd values of $\alpha$; this can be probably attributed to some symmetry of this particular case, and most likely does not extend to higher order calculations.

Let us end by mentioning that the decay rate, derived from the imaginary part of 
the self-energy, has a secular character, as expected. Even small decay rates 
could  give an important effect when integrated over large periods of time. 
Moreover, dissipation is a generic phenomenon which appears even in the vacuum.
 Therefore, if one considers remote times or large time lapses, 
the dissipative processes will become relevant, specially in situations in which the expansion rate, or the curvature, are large.

\acknowledgments{Many of the results presented in this paper originated from a longstanding research collaboration with Renaud Parentani and Enric Verdaguer,  triggered by our common attendance to the Peyresq meetings.  I am very grateful to both of them. This work is also partially supported by the Research Projects MEC FPA2007-66665C02-02 and DURSI 2005SGR-00082. }

\bibliography{books,articles}

\end{document}